# Non-invasive force measurement reveals the number of active kinesins on a synaptic vesicle precursor in axonal transport regulated by ARL-8

## Authors
Kumiko Hayashi[a]*, Shin Hasegawa[a], Takashi Sagawa[b], Sohei Tasaki[c] and Shinsuke Niwa[d]*

[a] Department of Applied Physics, Graduate School of Engineering, Tohoku University, Sendai, Japan

[b] Advanced ICT Research Institute, National Institute of Information and Communications Technology, Kobe, Japan

[c] Frontier Research Institute for Interdisciplinary Sciences (FRIS) and Graduate School of Sciences, Tohoku University, Sendai, Japan

[d] Frontier Research Institute for Interdisciplinary Sciences (FRIS) and Graduate School of Life Sciences, Tohoku University, Sendai, Japan

*Correspondence to:
kumiko@camp.apph.tohoku.ac.jp, shinsuke.niwa.c8@tohoku.ac.jp

**Abstract**

Kinesin superfamily protein UNC-104, a member of the kinesin-3 family, transports synaptic vesicle precursors (SVPs). In this study, the number of active UNC-104 molecules hauling a single SVP in axons in the worm *Caenorhabditis elegans* was counted by applying a newly developed non-invasive force measurement technique. The distribution of the force acting on a SVP transported by UNC-104 was spread out over several clusters, implying the presence of several force-producing units (FPUs). We then compared the number of FPUs in the wild-type worms with that in *arl-8* gene-deletion mutant worms. ARL-8 is a SVP-bound arf-like small guanosine triphosphatase, and is known to promote unlocking of the autoinhibition of the motor, which is critical for avoiding unnecessary consumption of adenosine triphosphate when the motor does not bind to a SVP. There were fewer FPUs in the *arl-8* mutant worms. This finding indicates that a lack of ARL-8 decreased the number of active UNC-104 motors, which then led to a decrease in the number of motors responsible for SVP transport.

**Introduction**

Due to the morphology of neurons, efficient communication between the cell body and distal processes requires fast cargo vesicle transport in the axon.[1, 2] Kinesin superfamily proteins and cytoplasmic dynein, which are adenosine triphosphate (ATP)-dependent molecular motors, haul cargo vesicles anterogradely toward the terminal, and retrogradely toward the cell body.[3-5] Neuronal communication largely depends on the delivery and receipt of synaptic vesicles in particular. The components of the synaptic vesicles are synthesized in the cell body and are then parcelled in synaptic vesicle precursors (SVPs). The SVPs are transported to the synapses, which are located around the axon terminal, by the kinesin superfamily protein UNC-104/KIF1A, which belongs to the kinesin-3 family. [6-8]

UNC-104/KIF1A was originally identified in the worm *Caenorhabditis elegans* (*C. elegans*).[7, 8] ARL-8, a molecule related to UNC-104, which is a SVP-bound arf-like small guanosine triphosphatase (GTPase), has recently been reported to activate the motor by releasing its autoinhibition.[9-11] The auto-inhibitory mechanism, which is a common mechanism for the regulation of molecular motors, is thought to be required to avoid both unnecessary cargo transport and futile ATP hydrolysis in cells when the motors do not bind to cargos. A lack of the regulatory protein ARL-8 leads to defects in axon-specific sorting of synaptic materials, decreased synapse density, and impaired cargo vesicle transport *in vivo*.[10]

In this paper, we investigated the physical mechanism underlying the finding that the absence of ARL-8 leads to decreased ability of UNC-104 to transport SVPs, as reported in a previous study.[10] Our hypothesis regarding this issue is summarized in Fig. 1. In our model, more motors can bind to a SVP in the presence of ARL-8 than in its absence (Fig. 1(b)), because ARL-8 promotes the activation of UNC-104 (Fig. 1(c)). This model is consistent with the observation that a basal level of axonal transport is observed even in loss-of-function mutants of *arl-8*.[10, 11] Note that only the activated form of UNC-104 can bind to a SVP, as has been shown previously.[10] The decreased number of motors hauling a SVP is thought to be the physical factor that lowers the SVP transport ability. For example, the run length of transported SVPs may become shorter as the number of motors decreases.

In order to verify our model, we investigated the number of motors hauling a single SVP by measuring the force, exerted by the motors, acting on a SVP. This force is thought to be a good indicator of the number of motors that carry a SVP, because a larger force acts on the SVP when it is transported by more motors (Fig.

1(b)). We then compared the results of wild-type worms with those of the *arl-8* mutant worms, in which *arl-8* was deleted.[10]

We have developed an approach for non-invasive force measurement based on the fluctuation theorem, which describes the relationship of non-equilibrium statistical mechanics regarding entropy production in non-equilibrium states.[12, 13] For the membrane vesicles transported by kinesin and dynein in *in vitro* cultured neurons, the force acting on a single vesicle has been estimated using this measurement method.[14] In this study, we also investigated the force acting on SVPs using the same force measurement method. First, we observed the directional motion of a GFP-labelled SVP in the axon of a *C. elegans* worm by using fluorescence microscopy, and obtained the center position of the SVP. The center position fluctuates due to thermal noise, collisions with other intercellular organelles, stochastic stepping motion of the motors caused by ATP hydrolysis, and other factors, while exhibiting the directional motion specified by the motors.

Second, we calculated the force indicator $\chi$ (eqn. (1)) from the fluctuating center position, which describes a quantity that is constructed based on the fluctuation theorem (see Methods section). The measurement was repeated many times for a large number of SVPs, to obtain the distribution of $\chi$. The distribution of $\chi$ was found to form several clusters, indicating the existence of multiple force-producing units (FPUs). From the calibrated force value of 1 FPU (about 5 pN), it was estimated that 1 FPU corresponds to a UNC-104 dimer.

Finally, we compared the distribution of the force indicator $\chi$ for the wild-type *C. elegans* with that for the *arl-8*-deleted mutant. As was expected from the model depicted in Fig. 1(b), we found that the number of FPUs was decreased in the mutant worm. The number of motors on a SVP remains poorly understood, yet this parameter is one of the important factors in understanding the physics of axonal transport; thus, our results are of general significance for the field of neuroscience.

**Materials and methods**
**Sample preparation**

*C. elegans* stains, *wyIs251*[Pmig-13::gfp::rab-3; Podr-1::gfp] and *arl-8*(*wy271*); *wyIs251* have been described in previous studies.[10, 11]

**Culture**

*C. elegans* was maintained on OP50 feeder bacteria on nematode agar plates (NGM) agar plates, as per the standard protocol.[15] Strains were maintained at 20ºC.

## Observation by fluorescence microscopy

A cover glass (32 mm × 24 mm, Matsunami Glass Ind., Ltd., Tokyo, Japan) was coated with 2% agar (Wako, Osaka, Japan). A volume of 20 µl of 25 mM levamisole mixed with 5 µl 200-nm-sized polystyrene beads (Polysciences Inc., Warrington, PA, USA) was dropped onto the cover glass. The polystyrene beads increased the friction, inhibiting the worms' movement[16] and levamisole paralyzed worms. Ten to 20 worms were moved from a culture dish to the medium on the cover glass. Another cover glass was placed over the first cover glass to confine the worms (forming a chamber) (Fig. 2(a)). The worms in the chamber were observed under a fluorescence microscope (IX83, Olympus, Tokyo, Japan) at room temperature (Fig. 2(b)). The images of a GFP-labelled SVP in a worm's neuron were obtained using a 150× objective lens (UApoN 150x/1.45, Olympus) and a sCMOS camera (OLCA-Flash4.0 V2, Hamamastu Photonics, Hamamatsu, Japan) at 100 frames per second. The center position of each SVP was determined from the recorded images using the custom-made software, Tyrant (see below). We focused on the displacement along the direction of the SVP's motion $X(t)$ (Fig. 2(c)). Data were collected from 40 vesicles from 33 individual wild-type worms in 21 different chambers, and from 40 vesicles from 36 individual *arl-8* mutant worms in 22 different chambers. The worms for observations were chosen randomly after visual inspection, and the trajectories that lasted about 0.5−1-s constant velocity runs were selected for analysis, as described in the main text. Note that although there were many SVPs in an axon, there were time intervals (rare events) in which single SVPs were tracked (Fig. S1). We collected all of the rare events that we could observe (n = 40 both for the wild-type and the mutant worms).

The accuracy of the position measurement was verified using beads (300-nm latex bead, Polysciences). The standard deviation (s.d.) of the position of the bead tightly attached to the glass surface was 8.3 ± 1.2 (s.d.) nm (based on four different beads). Because the standard deviation of the several-micron-sized synapses, which were fixed in worms, was 4.6 ± 1.7 nm (based on five different worms), the worms' bodies were considered to be tightly fixed during observations.

## Image analysis by Tyrant

Tyrant is a custom-developed software written in LabVIEW 2013 (National Instruments, Austin, TX, USA); this software analyses the trajectory of the center position of a SVP from the captured fluorescent images. To obtain the one-dimensional displacement of the SVP, all captured images were rotated so that the moving direction of SVP was parallel to the *x*-axis in the images. First, noise in a fluorescent image was removed using a 3 × 3 median filter. Next, the fluorescent spot of a SVP in the filtered image was cropped by an appropriately sized box (red boxes in Fig. 2(b)). The fluorescent intensity of the spot in the cropped image was fitted with a two-dimensional Gaussian function to calculate the center position of the spot. This procedure was repeated for each selected SVP for a specified time interval.

**Analysis of fluctuation using the fluctuation theorem**

The force indicator $\chi$, which was first introduced in our previous study,[14] is defined by

$$\chi = \ln[P(\Delta X)/ P(-\Delta X)]/\Delta X \tag{1}$$

from the distribution $P(\Delta X)$ of the displacement $\Delta X = X(t + \Delta t) - X(t)$ (e.g., Fig. 2(d), inset) in the constant velocity segment (e.g., Fig. 2(d)). $P(\Delta X)$ was fitted with a Gaussian function

$$P(\Delta X) = \exp(-(\Delta X - b)^2/2a)/(2\pi a)^{0.5} \tag{2}$$

where the fitting parameters $a$ and $b$ correspond to the variance and the mean of the distribution (e.g., Fig. 2(e)). By substituting eqn. (2) into eqn. (1),

$$\chi = 2b/a \tag{3}$$

Thus, $\chi$ was calculated as $2b/a$ for each $P(\Delta X)$ for various intervals $\Delta t$ from 10 ms to 100 ms (e.g., Fig. 2(f)). The converged value ($\chi^*$) (see Fig. 2(f) for example) of $\chi$ was related to a drag force ($F$) acting on a cargo by the relation[14]

$$F = k_B T_{eff} \chi^* \tag{4}$$

where $k_B$ is the Boltzmann constant and $T_{eff}$ is the effective temperature, which is a generalized temperature in a non-equilibrium system. It was suggested experimentally in a previous paper[14] that

$$F \propto \chi* \tag{5}$$

by measuring both $\chi^*$ and $F$ ($=\Gamma v$) for the same cargo vesicles, which exhibited both the constant velocity region and long pause; from the latter, the friction coefficient $\Gamma$ of a cargo vesicle can be estimated by using the power spectrum density of $X(t)$ in the long pause [14].

The error of $\chi$ was estimated based on the idea of the bootstrapping method. For the 10 different randomly selected segments in the constant velocity segments of $X(t)$ (the segments were half the length of the original), we calculated the values of χ (e.g., 10 thin curves in Fig. 2(f)). The standard deviation of $\chi^*$ was estimated to be 10%.

**Smoothing filtering and affinity propagation**

A smoothing filter was applied to the values of $\chi$ to reduce variation in the raw data of $\chi$ as a function of $\Delta t$ (Fig. 2 (f)). Here, we used the averaging filter $\chi^f(\Delta t) = (\chi(\Delta t - 10\text{ms}) + \chi(\Delta t) + \chi(\Delta t + 10\text{ms}))/3$, which is one of the simplest filters. Note that, in this paper, the converged value $\chi^*$ was defined as $\chi^* = \chi f(\Delta t = 80 \text{ ms})$.

Affinity propagation,[17,18] an exemplar-based clustering method that does not require the number of clusters, was then adopted for clustering the smoothing-filtered two-dimensional data ($\chi^*$, $\chi_m$) where $\chi_m$ is the mean value of $\chi$ from $\Delta t =$ 10 ms to $\Delta t =$ 100 ms. The method was applied by using "APCluster" implemented in the R software[18]. The clustering was stable for the wide range of parameters used in the method (Fig. S2).

**Force calibration**

In order to obtain the force value ($F$) from $\chi^*$ through the relation (eqn. (4)), we estimated the conversion factor

$$k_B T_{\text{eff}} = \Gamma v/\chi^* \qquad (6)$$

where $F = \Gamma v = \chi^* k_B T_{\text{eff}}$. When eqn. (5) is valid[14], the conversion factor from $\chi^*$ to $F$ is common for all SVPs. Accordingly, we used $\Gamma_a v_a/\chi^*_a$ as the conversion factor, where the index represents the averaged value of each quantity. The averaged value of the friction coefficient $\Gamma$ of the paused SVPs on microtubules was estimated to be 0.0044 ± 0.0024 pN s/nm (n = 6), which was calculated in the same way as reference 14 as follows. First, the position of the SVP was measured (Fig. S3(a)). Next, the power spectrum and distribution of the position along microtubules were calculated (Fig. S3(b)). Then, from fitting of the power spectrum of the position of the SVP by a Lorentzian function, $\Gamma/k$ was estimated, where $k$ is a stiffness of the pausing SVPs. This $k$ could be measured from the fitting of the distribution by the Boltzmann distribution $P_B(X) = (k/2\pi k_B T)^{0.5} \exp(-k(X - X_a)^2/2k_B T)$, where $T$ (= 25°C) is the room temperature (Fig. S3(c)). Because $\Gamma$ thus measured had a large error, which was the limitation of our *in vivo* experiments, this calibration provided only a rough estimate of force from the measured $\chi^*$. For the wild-type worms, the averaged value of the velocity ($v_a$) of the constant velocity region (n = 40) was 2.2 μm/s and $\chi^*_a$ was 0.11 nm$^{-1}$ in our experiment. Then the conversion factor $\Gamma_a v_a / \chi^*_a$ was estimated to be 85 pNnm. Note that this calibrated force has a large error, because the standard deviation of $\Gamma_a$ was 50%.

## Results
### Non-invasive force measurement using the force indicator

We observed the axonal transport of SVPs in wild-type worms by fluorescence microscopy (Fig. 2(a), Supporting Movies). The worms were anaesthetized and fixed between cover glasses with highly viscous media (high density 200-nm-sized polystyrene beads) in order to fix the worms' bodies and to minimize the fluctuation resulting from their body movements (see Methods section). Many studies have characterized the morphology and properties of the DA9 neuron and have investigated axonal transport mechanisms using this system.[9-11, 19, 20] We therefore used the DA9 neuron as a model in this study. SVPs in the DA9 neuron were labelled with GFP::RAB-3 using a mig-13 promoter. The motion of each vesicle transported by the motors was tracked by using in-house software (Fig. 2(b), Methods). Transport continued without pauses for a few seconds. The direction of motion was set to be the $x$-direction (Fig. 2(c)). Then, the center position ($X(t)$) of a vesicle along an axon was obtained as a function of time ($t$) for anterograde transport (Fig. 2(d)). Note that we only focused on anterograde transport to investigate UNC-104, given that retrograde transport was a rare event and we could therefore not obtain enough data for this type of transport.

Our analysis target was the constant velocity segment of a transported SVP that lasted about 0.5−1 s (Fig. 2(d)). It was difficult to track the same vesicle in the DA9 axon for longer than 1 s, because the axoplasm was crowded with SVPs that were moving fast (ca. 3 µm/s) and disappeared from the screen within several seconds. When the recording rate was high enough (100 frames per second), fluctuating behaviour of the vesicle was observed even while it exhibited directional motion (inset of Fig. 2(d)). The origin of the fluctuation in the transported SVPs was considered to be mainly thermal noise, stochastic ATP hydrolysis of motors, and collision of the SVP with other vesicles and cytoskeletons. The fluctuation resulting from the worm's body movement was not regarded as a major factor, because we only observed the vesicles for the worms whose movements were highly suppressed by the high viscosity media as well as the anaesthesia (see Methods section).

The fluctuating behaviour of a SVP was represented by $\Delta X$, which is defined by $\Delta X = X(t + \Delta t) - X(t)$ (inset of Fig. 2(d)). Fig. 2(e) shows the probability distribution ($P(\Delta X)$) of $\Delta X$ for the case $\Delta t = 100$ ms. It was fitted well by a Gaussian function. Using $P(\Delta X)$, which was calculated for each $\Delta t$ (($0 \leq \Delta t \leq 100$ ms), we then calculated the force indicator $\chi$ (eqn. (1) in the Methods section) (thick black line in Fig. 2(f)). Applying the smoothing filter (the orange curve in Fig. 2(f), Method), it can be clearly seen that $\chi$ reaches a constant value $\chi^*$ for a large $\Delta t$ after a certain amount of relaxation time (see Supplementary Theory section in reference 14 regarding the relaxation time). This converged value $\chi^*$ is found to be proportional to the force ($F$) acting on a single vesicle, exerted by the motors during constant velocity motion (eqn. (5)). Practically, $\chi^*$ was determined by $\chi^* = (\chi\,(\Delta t = 70$ ms$) + \chi\,(\Delta t = 80$ ms$) + \chi\,(\Delta t = 90$ ms$)) / 3$ in this paper. Note that the error in $\chi$ was estimated to be 10% based on a bootstrapping method (thin lines in Fig. 2(f), (n = 10), see Methods section), which was consistent with the systematic error of the experimental system.

**Distribution of the force indicator χ for anterograde transport of a SVP**

For anterograde transport of SVPs, we investigated 40 vesicles from 33 individual wild-type worms (Fig. 3(a)). After applying the smoothing filter (Fig. 3(b)) and the affinity propagation analysis (Fig. 3(c)), we found that the data were distributed into four clusters (see Methods section for the filtering and analysis). Because the converged value $\chi^*$ is proportional to the force ($F$) acting on a vesicle (eqn. (5)), this discrete behaviour of $\chi$ indicates the existence of four FPUs in SVP anterograde transport.

Next, we also investigated 40 vesicles from 36 different *arl-8*-deleted mutant worms (Fig. 4(a)). Similarly, after applying the smoothing filter (Fig. 4(b)) and the affinity propagation analysis (Fig. 4(c)), we found that the data were distributed into three clusters. The numbers of FPUs are summarized in Fig. 5. Compared with the case of the wild-type (Fig. 5(a)), the proportion of 3 FPUs was decreased, while the proportion of 1 FPUs was increased in the *arl-8*-deleted mutants (Fig. 5(b)). These evaluations thus indicated that the number of FPUs that hauled a SVP was reduced by 20% in the absence of ARL-8, as compared to the number of motors in wild-type worms.

**Force-velocity relation for the SVP anterograde transport**

The force-velocity relation for anterograde transport of SVPs was also investigated. By using the relation $F \propto \chi^*$, instead of force ($F$), $\chi^*$ was plotted as a function of velocity ($v$) for the wild-type worms (Fig. 6(a), filled symbols) and the *arl-8* mutant worms (Fig. 6(a), open symbols). Here, $v$ was measured as the slope of the constant velocity region used to calculate $\chi$ (e.g., Fig. 2(d)). It is seen that the mean velocity slightly increases as the number of FPUs increases in Fig. 6(a). The increase resulted from the reduction in load for a single motor when transport involved multiple motors.

The conversion factor (eqn. (6)) from $\chi^*$ to $F$ was roughly estimated as $\Gamma_a v_a / \chi_a^*$ (the index "a" represents the averaged value of each quantity), where $\Gamma$ is the friction coefficient of a SVP along microtubules. The calibrated force obtained from this rough estimation (see Method section) is shown in the superior axis in Fig. 6(a). Here, 1 FPU was considered to be a UNC-104 dimer, because the calibrated force value (about 5 pN) of 1 FPU was similar to the stall force value of the dimer measured in *in vitro* single-molecule experiments.[23]

Lastly, we compared the $\chi^*-v$ relation of the wild-type worms (filled symbols) with that of the *arl-8*-deleted mutant worms (open symbols). Their similar $\chi^*-v$ relation suggests that the force generation mechanism of UNC-104 motors was not largely affected by the lack of ARL-8.

**Fluorescence intensity of SVPs**

In Fig. 6(b), the fluorescence intensity (FI) of SVPs is plotted as a function of $\chi^*$ for the wild-type worms (filled symbols) and the *arl-8* mutant worms (open symbols). $FI^{1/2}$ was considered to represent the radius of a SVP. Because the correlation coefficient ($R^2$) was 0.40, there was a weak correlation between $FI^{1/2}$ and $\chi^*$. This implies that a large SVP tends to be conveyed by multiple motors because many motors can be attached to a large SVP (Fig. 6(c)).

**Discussion**

In a previous study [10], it was reported that absence of *arl-8* causes mislocalization of synaptic vesicles in the DA9 neuron of *C. elegans*, implying that the lack of ARL-8 reduced the anterograde transport capability of SVPs. However, the physical parameters in the vesicle transport that are affected by the lack of ARL-8 remained elusive. In this study, we identified one of these physical parameters, i.e., the number of the motors carrying a SVP. We assumed that the binding of UNC-104 to a SVP is promoted by the presence of ARL-8, because ARL-8 promoted the chemical transition from the auto-inhibited state to the uninhibited state of UNC-104 (Fig. 1(c)), and only UNC-104 in the uninhibited state can bind to a SVP. Consequently, the number of motors carrying a single SVP was increased in the presence of ARL-8. In order to verify our assumption, we investigated the force acting on a SVP by counting the number of motors, because the force increases directly with the number of motors attached to the SVP (Fig. 1(b)). Then, using a non-invasive force measurement method based on the fluctuation theorem of non-equilibrium statistical mechanics,[14] the force indicator $\chi$ (eqn. (1)) was calculated for a transported SVP (Fig. 2(f)). The distribution of $\chi$ clustered into several groups (Fig. 3), implying the existence of multiple FPUs. In the case of the *arl-8*-deleted mutant, the number of clusters in the distribution of $\chi$ was decreased (Fig. 4). Thus, we concluded that the number of UNC-104 motors hauling a single SVP was decreased in the absence of ARL-8 (Fig. 5). Below, we discuss several issues related to these main findings.

### Quantal behaviour of force indicator $\chi$

It was investigated by using optical tweezers in living cells that the stall force acting on a single cargo is quantal, reflecting the number of motors hauling the cargo.[21, 22] This quantal behaviour is one of typical properties of force in intracellular cargo transport systems. Similarly, in this paper, discrete behaviour was observed in the distribution of the force indicator $\chi$ in neuronal SVP transport (Figs. 3 and 4). The observation of the quantal behaviour of $\chi$ supported the relation $F \propto \chi^*$ proposed in our previous paper.[14]

### Run length

It has been shown by *in vitro* single-molecule experiments that the run length of kinesin motors increased with the number of motors, by controlling the number of DNA scaffolds.[24] Using these reported values of the run length and the number of FPUs in Fig. 5, the averaged run length of a SVP for the *arl-8*-deleted mutant worms was calculated to be 70% shorter than that of the wild-type worms. Because it has been reported that the distance of synapses from the cell body was reduced by about 60% by the deletion of *arl-8*,[10] the decreased run length in the mutant may contribute to induction of synapse misplacement in axons.

### Conclusions

The fluctuation theorem of non-equilibrium statistical mechanics[12, 13] has been experimentally applied to many non-equilibrium physical systems, such as colloidal particle systems,[25] granular systems,[26] and turbulence systems,[27] as well as single-molecule experiments.[28, 29] In this paper, we applied the theorem to axonal transport in living worms. While *in vitro* single-molecule experiments on kinesin and dynein motors have revealed the basic dynamic properties of motors accompanied by ATP hydrolysis, their intracellular properties remain unknown. In particular, cooperative transport by multiple motors is one of the properties that differ from those observed in *in vitro* single-molecule experiments. Since physical measurements *in vivo* is difficult because of the complexity of the cell environments, non-invasive passive measurement using fluctuation properties are powerful tools for evaluating intracellular physical quantities. The non-invasive force measurement approach proposed here, which is based on the fluctuation theorem, is likely to find application in a wide range of complex biological *in vivo* systems.

## Conflicts of interest

There are no conflicts to declare.

## Acknowledgements

We thank Dr. Y. Okada for assistance with the initial planning of the study; Mr. Y. Saito, and Ms. S. Sato for help with the experiments; and the members belonging to the Innovative Areas "Soft Molecular Systems" and the young scientists of FRIS, Tohoku University for discussions about the study. This work was supported by PRIME, AMED and Grant-in-Aid for Scientific Research (KAKENHI) (grant numbers 26104501, 26115702, 26310204, and 16H00819), the Ministry of Education, Culture, Sports, Science, and Technology to K. H., as well as by the grant from Daiichi Sankyo Foundation of Life Science and Grant-in-Aid for Scientific Research (KAKENHI) (grant numbers 17H0510), the Ministry of Education, Culture, Sports, Science, and Technology to S. N..

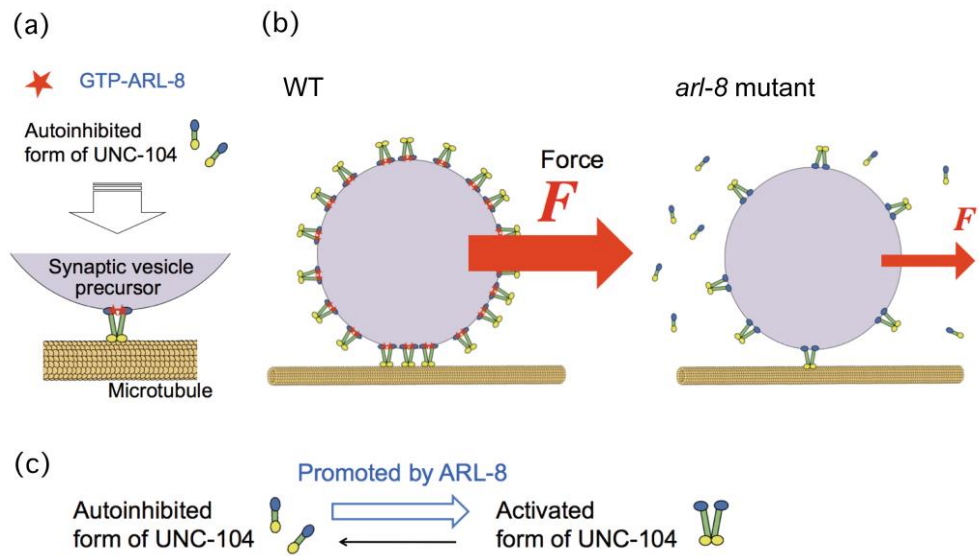

**Fig. 1** Schematics of our model. (a) Mechanism of auto-inhibition of UNC-104 released by ARL-8 binding, as suggested in a previous paper.[10] Activation of UNC-104 motors is promoted by ARL-8 (red), and these then bind to a synaptic vesicle precursor (SVP). UNC-104 motors (yellow) carry the SVP (violet) along a microtubule (orange). (b) This model suggests that the lack of ARL-8 reduces the number of UNC-104 motors attached to a SVP. Consequently, the number of UNC-104 motors in charge of SVP transport is also reduced. Because the force ($F$) generated by the motors acting on a SVP increases directly with the number of motors, the force can be used for estimating the motor number. (c) The transition from the auto-inhibited state of UNC-104 to its activated state is promoted by ARL-8.

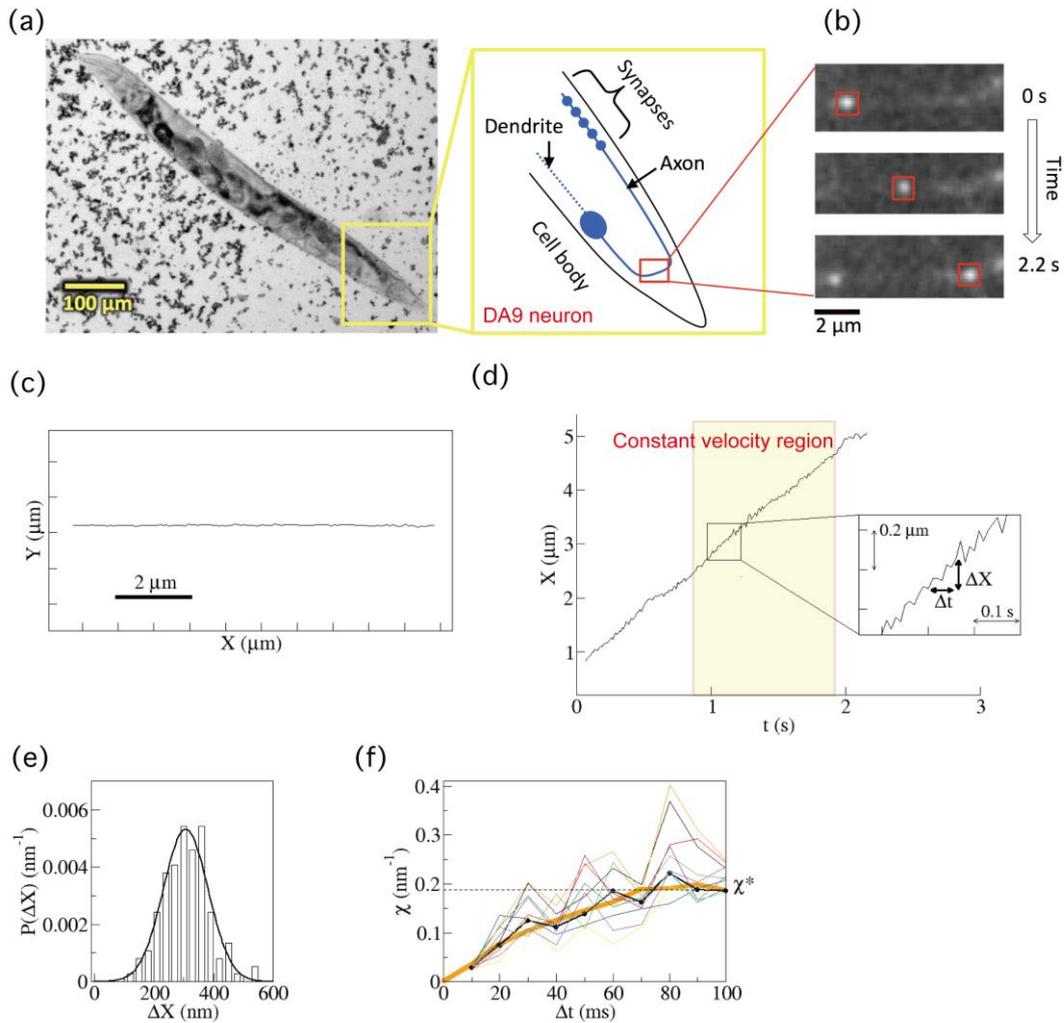

**Fig. 2** Fluorescence observation of a synaptic vesicle precursor (SVP) and calculation of the force indicator $\chi$. (a) A bright-field micrograph of a *C. elegans* worm whose body was fixed in a chamber by the high viscosity medium (high-density 200-nm-sized polystyrene beads) and anaesthesia (see Methods section). SVPs were observed in the axon of the DA9 neuron. (b) Time-series fluorescence images of a SVP. The motion was tracked by in-house software (see Methods section). (c) An example of a 2D-trajectory of the movement of a SVP. The direction of the movement was set to be the *x*-direction. (d) Center position ($X$) of a SVP for anterograde transport as a function of time. Our analysis target was a constant velocity region that lasted longer than 0.5 s. When the recorded speed was high enough (100 frames per second), the fluctuation of $X$ was observed (inset). Here, $\Delta X = X(t + \Delta t) - X(t)$. (e) Probability distribution ($P(\Delta X)$) of $\Delta X$ for the cases of 100 ms. The distribution was well fitted by a Gaussian function (thick curve). (f) An example of the force indicator $\chi$ (eqn. (1)) calculated for each $P(\Delta X)$

at a different $\Delta t$ (10 ms $\leq \Delta t \leq$ 100 ms) (black). After a relaxation time, $\chi$ reached a constant value $\chi^*$ (see Supplementary Material of reference 14 for the relaxation time). The thin curves (n = 10) represent $\chi$ calculated based on the boot-strapping method for the error estimation (see Methods section). The estimated error of $\chi$ was 10%, which was consistent with the systematic error of the experimental system. The thick orange curve represents $\chi$ after the smoothing filtering (see Methods section).

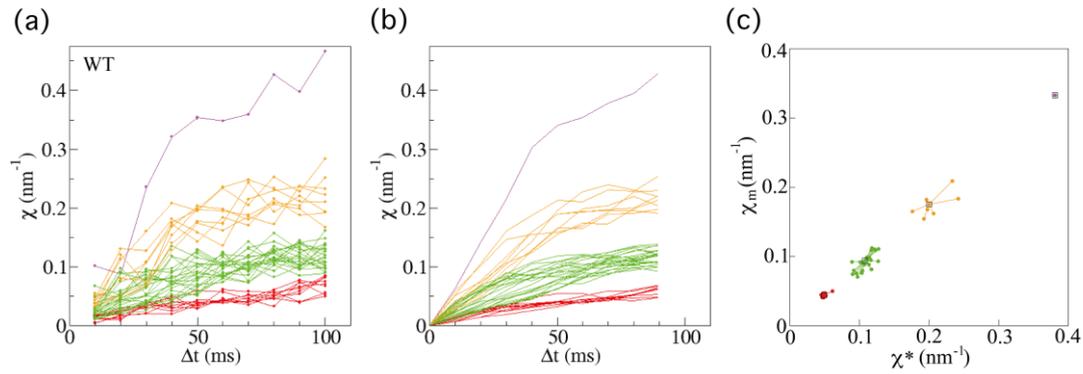

**Fig. 3** Force indicator $\chi$ for anterograde synaptic vesicle precursor (SVP) transport in wild-type *C. elegans*. (a) $\chi$ (eqn. (1)) as a function of $\Delta t$ for SVP anterograde runs for wild-type (WT) *C. elegans* worms (n = 40 SVPs from 33 different worms). (b) Results of applying the smoothing filter (see Methods section) to the data in Fig. 3(a). (c) The affinity propagation analysis (see Methods section) applied to the data in Fig. 3(b). There are four clusters in the case of wild-type worms. Each color denotes a cluster, representing different force-producing units.

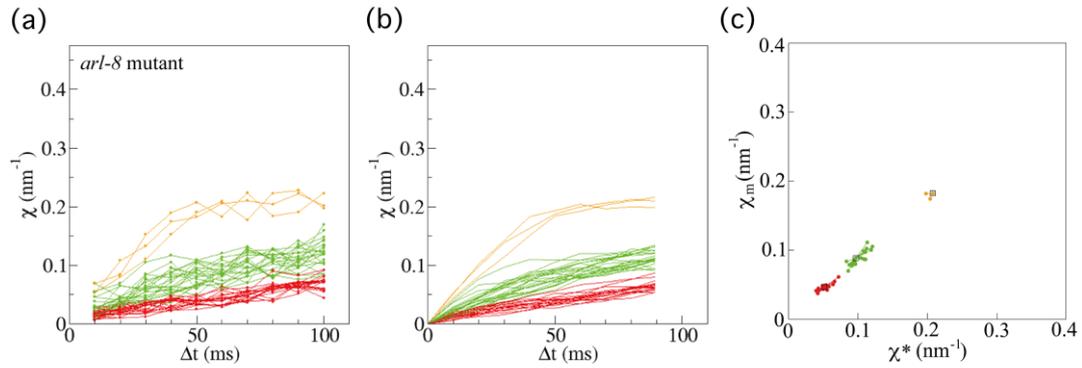

**Fig. 4** Force indicator $\chi$ for synaptic vesicle precursor (SVP) anterograde transport for the *arl-8*-deleted mutant. (a) $\chi$ (eqn. (1)) as a function of $\Delta t$ for SVP anterograde runs for the *arl-8*-deleted mutant *C. elegans* worms ($n$ = 40 SVPs from 36 different worms). (b) Results of applying the smoothing filter (see Method section) to the data in Fig. 4(a). (c) The affinity propagation analysis (see Method section) applied to the data in Fig. 4(b). There are three clusters in the case of the *arl-8*-deleted mutant worms. Each color denotes a cluster, representing different force-producing units (FPUs).

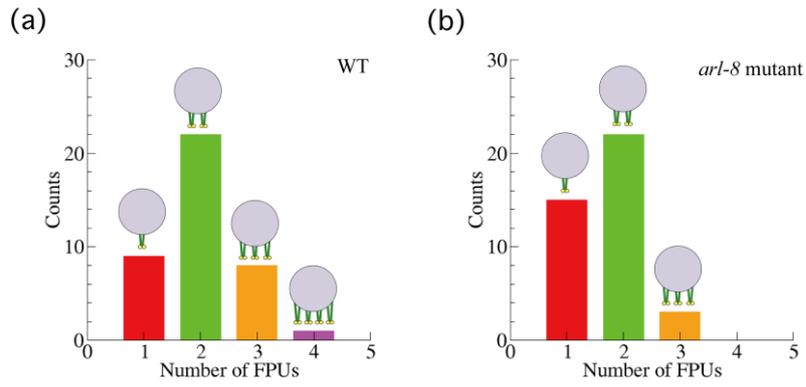

**Fig. 5** Number of force-producing units (FPUs). The number of FPUs involved in Figs. 3 and 4, in the case of the wild-type (WT) worms (a), and in the case of the *arl-8*-deleted mutant worms (b).

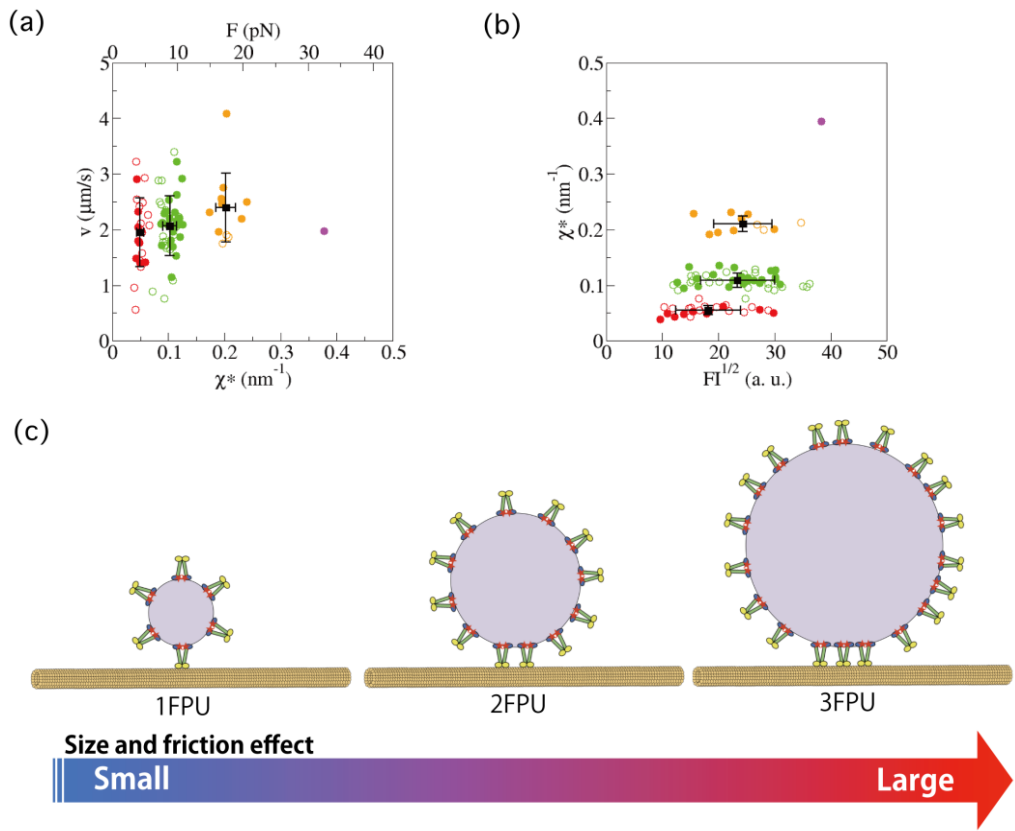

**Fig. 6** (a) Force-velocity relation for anterograde synaptic vesicle precursor (SVP) transport. By using the relation $F \propto \chi^*$, instead of force ($F$), $\chi^*$ was investigated as a function of velocity ($v$) for the wild-type (WT) worms (filled circles) and the *arl-8* mutant worms (open circles). Here, $v$ was measured as the slope of the constant velocity region used to calculate $\chi$ (e.g., Fig. 2(d)). Each color corresponds to a group shown in Figs. 3 and 4. In the superior axis, the calibrated force value is represented (see Methods). The error bars represents the standard deviations. (b) $\chi^*$ as a function of the square root of the fluorescence intensity (FI) of a SVP. $FI^{1/2}$ corresponds to the radius of a SVP. The error bars represents the standard deviations. There was a weak correlation between $FI^{1/2}$ and $\chi^*$ ($R^2=0.40$). (c) Schematics explaining the graphs in Fig. 6(b). Each element of this schematics (e.g., UNC-104, microtubules, etc.) is explained in Fig. 1(a).

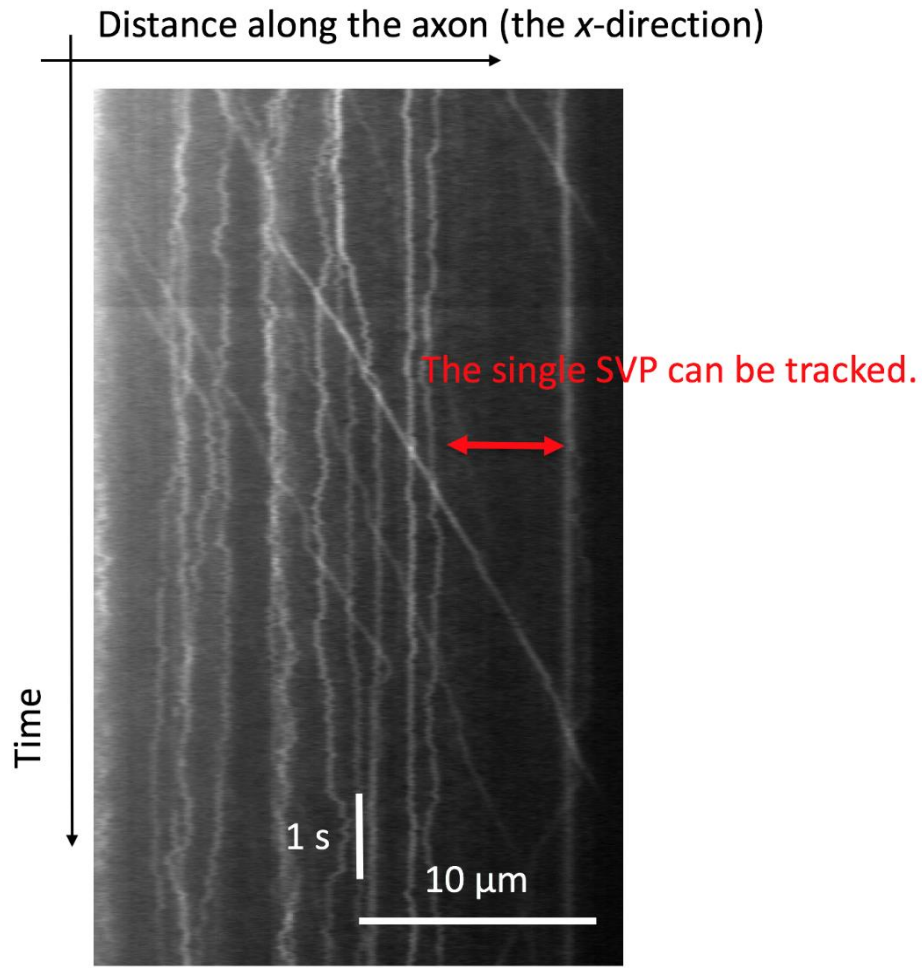

**Fig. S1 Kymograph for SVPs.** There were many SVPs in axons. The kymograph represents the motion of SVPs along the direction of the axon as a function of time. Each bright line represents the motion of a SVP. Several SVPs moved anterogradely. There were time intervals that the single SVP could be tracked (*e.g.*, the red arrows) even though the axon was clouded with many SVPs. We investigated such rare time intervals.

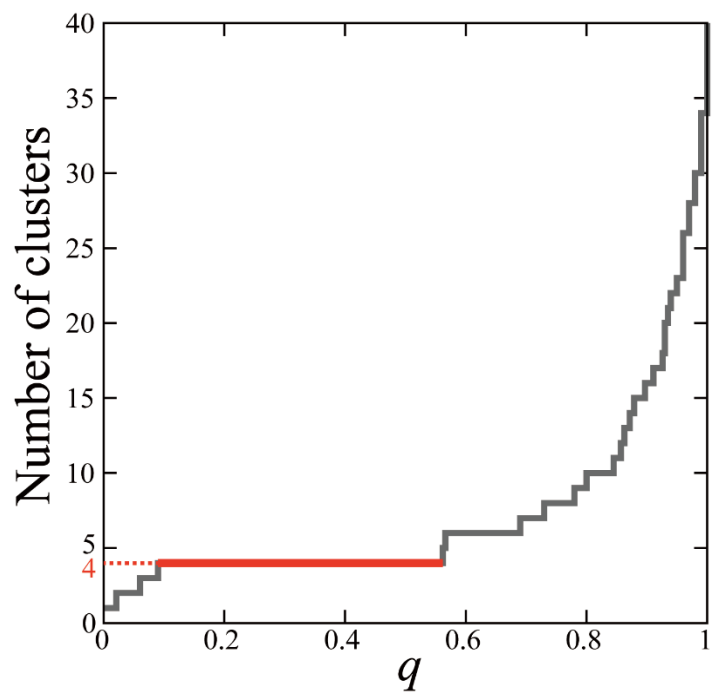

**Fig. S2 Stability of the clustering analysis (Affinity propagation).** The effect of the parameter $q$, which controls the input preference of the affinity propagation method[18], on the result (the number of clusters). It is known that the value $q=0.5$ results in a modulate number of clusters in general[18]. It was seen that the cluster number was stable for the wide range of $q$ for our experimental data (Fig. 3).

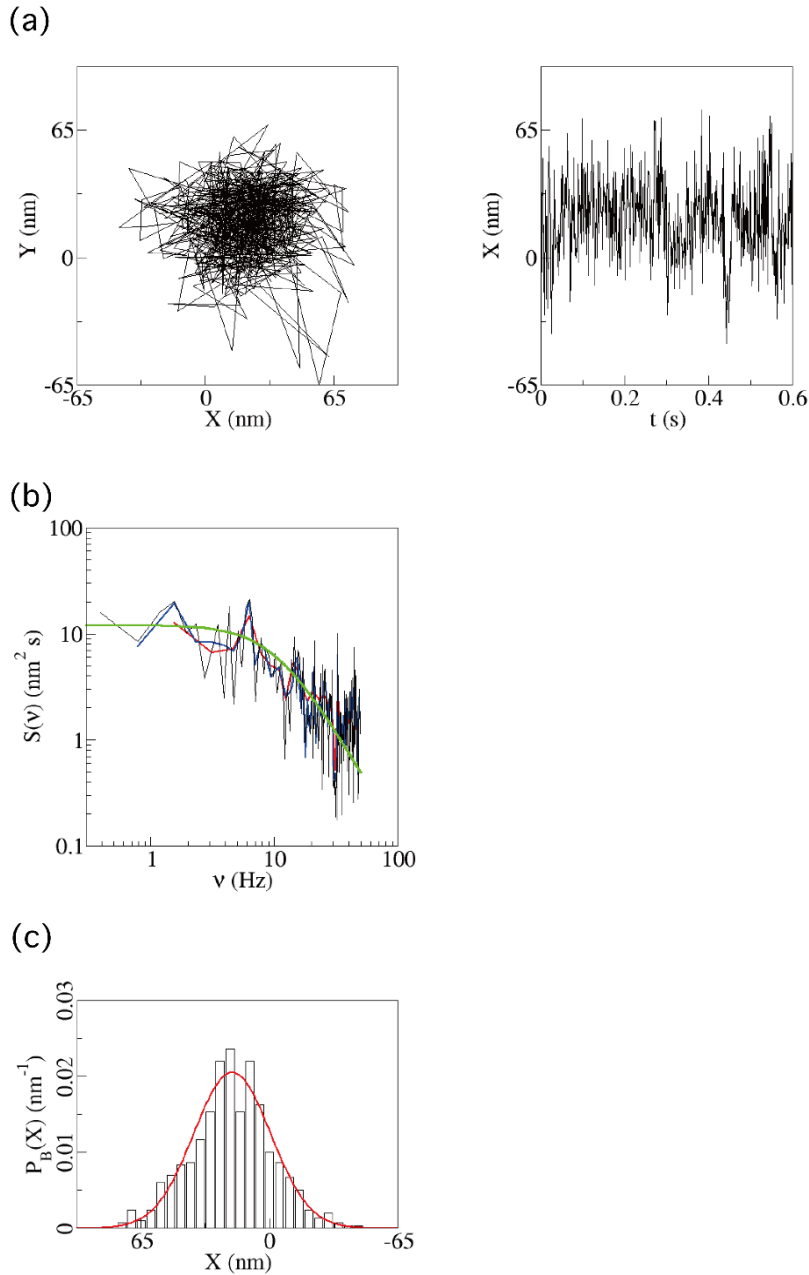

**Fig. S3 Force calibration.** (a) The position ($X$, $Y$) of a paused synaptic vesicle precursor (SVP) (left). The *x*-direction is the direction of motion of transported SVPs. $X(t)$ as a function of time is plotted (right). (b) The power spectrum of the position $X(t)$ in the cases $N_w$=64 (red), 128 (blue) and 256 (black), respectively ($N_w$ is the window of the Fourier transform). It was fitted by the Lorenztian $c/(1+(\Gamma/k\cdot\nu)^2)$ (green), where $c$ is a constant. (c) The probability distribution, $P_B(X)$, of $X$. It wa fitted by the Boltzmann distribution $P_B(X)=(k/2\pi k_B T)^{0.5}\exp(-k\,(X-X_a)^2/2k_B T)$.